\documentclass[showpacs,showkeys,preprintnumbers, amsmath,amssymb,preprint]{revtex4-1}
\usepackage{amssymb}
\usepackage{amsfonts}
\usepackage{amsmath}
\usepackage{graphicx}
\usepackage{dcolumn}
\usepackage{bm}
\usepackage{makeidx}
\usepackage[latin1]{inputenc}
\usepackage[english, english]{babel}
\usepackage[dvips]{hyperref}

\usepackage{graphicx}
\def\be{\begin{equation}}
 \def\ee{\end{equation}}
 \def\bea{\begin{eqnarray}}
 \def\eea{\end{eqnarray}}

\makeindex

\begin{document}

\title{Quasinormal modes
of Four Dimensional Topological Nonlinear Charged Lifshitz Black Holes}
\author{Ram\'{o}n B\'{e}car}
\email{rbecar@uct.cl}
\affiliation{Departamento de Ciencias Matem\'{a}ticas y F\'{\i}sicas, Universidad Cat\'{o}%
lica de Temuco, Montt 56, Casilla 15-D, Temuco, Chile}
\author{P. A. Gonz\'{a}lez}
\email{pablo.gonzalez@udp.cl}
\affiliation{Facultad de Ingenier\'{\i}a, Universidad Diego Portales, Avenida Ej\'{e}%
rcito Libertador 441, Casilla 298-V, Santiago, Chile.}
\author{Yerko V\'{a}squez.}
\email{yvasquez@userena.cl}
\affiliation{Departamento de F\'{\i}sica, Facultad de Ciencias, Universidad de La Serena,\\
Avenida Cisternas 1200, La Serena, Chile.}
\date{\today}

\begin{abstract}
We study scalar perturbations of four dimensional topological nonlinear charged Lifshitz black holes with spherical and plane transverse sections, and we find numerically the quasinormal
modes for scalar fields.
Then, we study the stability of these black holes under massive and massless scalar field perturbations. We focus our study on the dependence of the dynamical exponent,  the nonlinear exponent, the  angular  momentum and the mass of the scalar field in the modes. It is found that the modes are overdamped depending strongly on the dynamical exponent and the angular momentum of the scalar field for a spherical transverse section. In constrast, for plane transverse sections the modes are always overdamped.  
\end{abstract}

\maketitle

\tableofcontents

\newpage

\section{Introduction}

The gauge/gravity duality contains interesting gravity theories. One of them is 
the known as Lifshitz gravity that can be dual to scale-invariant field theories, being not conformally invariant. In this context, interesting properties are found when one generalizes the gauge/gravity duality to non-relativistic situations \cite{Kachru:2008yh,Hartnoll:2009ns, Brynjolfsson:2009ct,  Sin:2009wi, Schaposnik:2012cr, Momeni:2012tw, Bu:2012zzb, Keranen:2012mx, Zhao:2013pva, Lu:2013tza, Tallarita:2014bga, Taylor:2008tg, Roychowdhury:2015cva}, being the Lifshitz holographic superconductor  one of the most well studied systems.
Such theories exhibit the anisotropic
scale invariance $t\rightarrow \chi ^{z}t$, $x\rightarrow \chi x$, with $z\neq
1$, where $z$ is the relative scale dimension of time and space. 
Systems with such behavior appear, for instance, in the
description of strongly correlated electrons. 

In this work, we consider a matter distribution outside the event horizon of
the topological nonlinear charged Lifshitz
black hole in $4$-dimensions with a spherical  and plane transverse section and dynamical exponent $z$  \cite{Zangeneh:2015uwa}.
The matter is parameterized by scalar fields minimally coupled to gravity. Then, we obtain numerically the quasinormal frequencies (QNFs) for scalar fields, by using the improved AIM \cite{Cho:2009cj}, which is an improved version of the method proposed in references \cite{Ciftci, Ciftci:2005xn} and it has been applied successful in the context of quasinormal modes (QNMs) for different black hole geometries (see for instance \cite{Cho:2009cj, Cho:2011sf, Catalan:2013eza, Catalan:2014ama, Zhang:2015jda, Barakat:2006ki, Sybesma:2015oha, Gonzalez:2015gla, Becar:2015gca}). Then, we study
their stability under scalar perturbations.  We focus our study on dependence of the dynamical exponent,  the nonlinear exponent, the momentum angular and the mass of the scalar field in obtaining of quasinormal frequencies overdamped.  Mainly, motivated by a recent work, where the authors have shown that for $d > z+1$, at zero momenta, the modes are non-overdamped, whereas for $d \leq z + 1$ the system is always overdamped \cite{Sybesma:2015oha}. Contrary to other Lifshitz black holes, where 
the QNFs show the absence of a real part \cite{CuadrosMelgar:2011up, Gonzalez:2012de, Gonzalez:2012xc,Myung:2012cb, Becar:2012bj,Giacomini:2012hg,Catalan:2014una} 

In the gravity side of the gauge/gravity duality the QNFs \cite{Regge:1957td, Zerilli:1971wd,
Zerilli:1970se, Kokkotas:1999bd, Nollert:1999ji, Konoplya:2011qq} gives
information about the stability of black holes under matter fields that
evolve perturbatively in their exterior region, these fields are considered as mere test fields, without backreaction over the spacetime itself. Also, QNMs have shown to be related to the area and entropy spectrum of black holes horizon. Besides, the QNFs determine how fast a thermal state in the
boundary theory will reach thermal equilibrium according to the gauge/gravity duality \cite{Maldacena:1997re}, where the relaxation time of a
thermal state is proportional to the inverse
of the smallest imaginary part of the QNFs of the dual gravity background, which was
established due to the QNFs of the black hole being related to the poles of
the retarded correlation function of the corresponding perturbations of the
dual conformal field theory \cite{Birmingham:2001pj}. 

The paper is organized as follows. In Sec. \ref{Background} we give a brief review of
the topological nonlinear charged Lifshitz black holes that we will consider as background. In Sec. \ref{QNM} we calculate the QNFs of scalar perturbations numerically by using the improved AIM. Finally, our conclusions are in Sec. \ref{conclusion}.

\section{Topological nonlinear charged Lifshitz black holes}
\label{Background}
The topological nonlinear charged Lifshitz black holes that we consider is solution of the Einstein-dilaton gravity in the presence of a power-law and two linear Maxwell electromagnetic fields \cite{Zangeneh:2015uwa}. The action is given by
\begin{equation}\label{action4d}
S=-\frac{1}{16\pi}\int_M d^4x\sqrt{-g}\left( R -2(\nabla\phi)^2-2\Lambda_4+(-e^{-2\lambda_1\phi}F)^p-\sum_{i=2}^3 e^{-2\lambda_i\phi}H_i\right)~,
\end{equation}
where $R$ is the Ricci scalar on manifold M, $\phi$ is the dilaton field, $\Lambda_4$ is the cosmological constant, $\lambda_1$ and $\lambda_i$ are constants. $F$ and $H_i$ are the Maxwell invariants of electromagnetic fields $F_{\mu\nu}=\partial_{[\mu}A_{\nu]}$ and $(H_i)_{\mu\nu}=\partial_{[\mu}(B_i)_{\nu]}$, where $A_\mu$ and $(B_i)_\mu$ are the electromagnetic potentials. 
The following metric is solution of the equations of motion of the theory defined by the action (\ref{action4d})
\begin{equation}
ds^2=-\frac{r^{2z}}{l^{2z}}f(r)dt^2+\frac{l^2}{r^2}\frac{dr^2}{f(r)}+r^2d\Omega_k^2~,
\end{equation}
where $d\Omega_k^2$ is the metric of the spatial 2-section, which can have positive $k=1$, negative $k=-1$
or zero curvature $k=0$, and
\begin{equation}
f(r)=1+\frac{kl^2}{r^2z^2}-\frac{m}{r^{z+2}}+\frac{q^{2p}}{r^{\Gamma_4+z+2}}~,
\end{equation}
if the constant $\Lambda_4$ is
\begin{equation}
\Lambda_4=-\frac{(z+1)(z+2)}{2l^2}~.
\end{equation}
The gauge field is given by
 \begin{equation}
 F_{rt}=\frac{q_{1}b^{2(z-1)}}{r^{\Gamma_{4}+1}}~,
 \end{equation}
 and the gauge potential by
 \begin{equation}
 A_{t}=-\frac{q_{1}b^{2(z-1)}}{\Gamma_{4}r^{\Gamma_{4}}}~,
 \end{equation}
 where
 \begin{equation}
 q^{2p}=\frac{(2p-1)b^{2(z-1)}}{2\Gamma_4 l^{-2p(z-1)-2}}(2q_1^2)^p~,
 \end{equation}
 \begin{equation}
 \Gamma_4=z-2+\frac{2}{(2p-1)}~,
 \end{equation}
being $q_1$ and $b$ constants. Also, in order to have a finite mass, $\Gamma_4$ should be positive, which imposes the following restrictions on $p$ and $z$:
 \begin{itemize}
 \item
 For $p<1/2$, $z-1>(3-2p)/(1-2p)$,
 \item
 For $1/2<p\leq 3/2$, all $z (\geq 1)$ values are allowed,
 \item
 For $p>3/2$, $z-1>(2p-3)/(2p-1)$.
 \end{itemize}

\section{Quasinormal modes}
\label{QNM}
\begin{figure}[h]
\begin{center}
\includegraphics[width=0.45\textwidth]{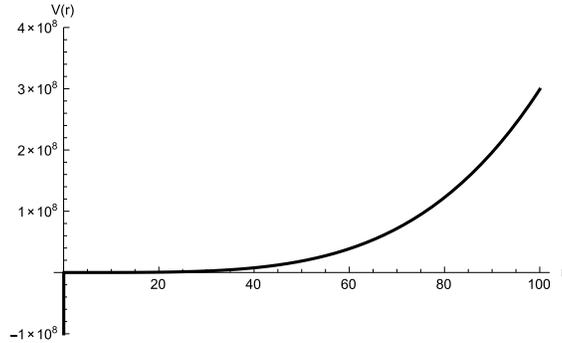}
\end{center}
\caption{The behavior of $V(r)$ with $l=1$, $m=1$, $q_1=0.1$, $m_s=0.1$, $b=1$, $z=2$, $p=2$ and $Q=2$.} \label{Potential1}
\end{figure}

\begin{figure}[h]
\begin{center}
\includegraphics[width=0.6\textwidth]{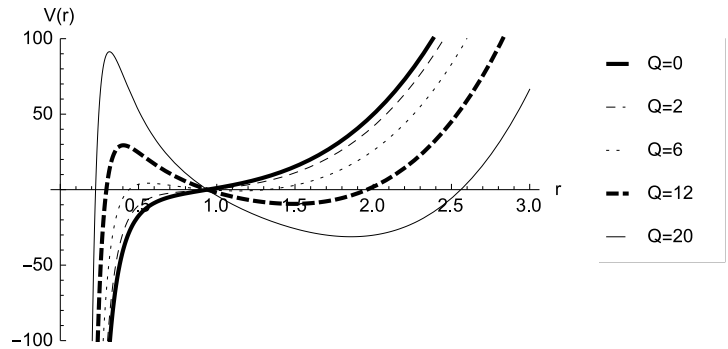}
\end{center}
\caption{The behavior of $V(r)$ with $l=1$, $m=1$, $q_1=0.1$, $m_s=0.1$, $b=1$, $z=2$, $p=2$ and $Q=0, 2, 6, 12, 20$.} \label{Potential2}
\end{figure}

The Klein-Gordon equation for a scalar field minimally coupled to curvature is
\begin{equation}
\frac{1}{\sqrt{-g}}\partial _{\mu }\left( \sqrt{-g}g^{\mu \nu }\partial
_{\nu }\right) \psi =m_s^{2}\psi ~,  \label{KGNM}
\end{equation}%
where $m_s$ is the mass of the scalar field $\psi $. Thus, the QNMs of scalar perturbations in the background of a four-dimensional topological nonlinear charged Lifshitz black holes are given by the scalar field solution of the Klein-Gordon equation with appropriate boundary conditions. 
Now, by means of the following ansatz
\begin{equation}
\psi =e^{-i\omega t}R(r)Y(\theta ,\phi )~,
\end{equation}%
where $Y(\theta ,\phi )$ is a normalizable harmonic function on the two-sphere
which satisfies the eigenvalues equation $\nabla^{2}Y(\theta ,\phi )=-QY(\theta ,\phi )$, where $Q=\ell(\ell+1)$ $\ell=0,1,2, ... $. Then,
the Klein-Gordon equation yields 
\begin{equation}
 \frac{1}{r^{z+3}}\partial_{r}\left[r^{z+3}f(r)\partial_{r}R\right]+\left[\left(\frac{l}{r}\right)^{2(z+1)}\frac{\omega^{2}}{f(r)}-\left(\frac{l}{r}\right)^{2}\frac{Q}{r^{2}}-\left(\frac{l}{r}\right)^{2}m_s^{2}\right] R(r)=0~. \label{radial}
\end{equation}%
Also, defining $R(r)$ as
 \begin{equation}
 R(r)=\frac{K(r)}{r}~,
 \end{equation}
and by using the tortoise coordinate $r_{\ast}$ given by
 \begin{equation}
 dr_{\ast}=\frac{l^{z+1}dr}{r^{z+1}f(r)}~,
 \end{equation}
 the Klein-Gordon equation can be written as a one-dimensional Schr\"{o}dinger equation
 \begin{equation}
 \frac{d^{2}K(r_{\ast})}{dr_{\ast}^{2}}+\left[\omega^{2}-V(r)\right] K(r_{\ast})=0~,
 \end{equation}
 where the effective potential $V(r)$ 
 \begin{equation}
 V(r)=\left(\frac{r}{l}\right)^{2z}f(r)\left(\frac{(z+1)}{l^2}f(r)+\frac{r}{l^2}f^{\prime}(r)-\frac{Q}{r^2}-m_s^{2}\right)~,
 \end{equation}
diverges at spatial infinity, see Fig. \ref{Potential1}. Therefore,
 we will consider that the field vanishes at the asymptotic region as boundary condition or Dirichlet boundary condition. In Fig. \ref{Potential2} we plot the behavior of the effective potential near the horizon for different values of $Q$. 
It is worth to mention that is not trivial to find analytical solutions to Eq. (\ref{radial}).
So, 
we will perform numerical studies
by using the improved AIM \cite{Cho:2009cj}. In order to implement the improved AIM
we make the following change of variables $u=1-r_H/r$ to Eq. (\ref{KGNM}). Then, the Klein-Gordon equation yields
\begin{eqnarray}
\nonumber && r_H^{z+1}(1-u)^{1-z}f(u)\partial^2_uR(u)+r_H^{z+1}(1-u)^{2}\left(\frac{f^{\prime}(u)}{(1-u)^{1+z}}+\frac{f(u)(1+z)}{(1-u)^{z+2}}\right)\partial_uR(u)\\
&&+\left( \frac{l^{2(z+1)}\omega^2(1-u)^{z-1}}{r_H^{z-1}f(u)}+\frac{r_H^{z-1}l^2Q}{(1-u)^{z-1}}-\frac{m_s^2r_H^{z+1}l^2}{(1-u)^{z+1}}\right)R(u)=0~.
\label{numericalmethod}
\end{eqnarray}
Now, in order to propose an ansatz for the scalar field, we must consider their behavior on the event horizon and at spatial infinity. Accordingly, on the horizon, $u\rightarrow 0$, their behavior is given by
\begin{equation}
R\left( u\rightarrow 0\right) \sim C_{1}u^{i\frac{l^{z+1}\omega}{r_H^z f^{\prime}(0)}}+C_{2}u^{-i\frac{l^{z+1}\omega}{r_H^z f^{\prime}(0)}}~,
\end{equation}
So, if we consider only ingoing waves on the horizon, we must impose $C_{1}=0$. Also, asymptotically, from Eq. (\ref{numericalmethod}), the scalar field behaves as
\begin{equation}\label{asymp}
R \left( u\rightarrow 1\right) \sim D_{1}\left( 1-u\right)
^{1/2(1+(1+z)-\sqrt{(2+z)^2+4m^2l^2})}+D_{2}\left( 1-u\right) ^{1/2(1+(1+z)+\sqrt{(2+z)^2+4m^2l^2})}~.
\end{equation}
So, in order to have a null field at infinity we must impose $D_{1}=0 $. Therefore, taking into account these behaviors we define
\begin{equation}
R\left( u\right) =u^{-i\frac{l^{z+1}\omega}{r_H^z f^{\prime}(0)}}\left( 1-u\right) ^{1/2(1+(1+z)+\sqrt{(2+z)^2+4m^2l^2})}
\end{equation}
as ansatz. Then, by inserting these fields in Eq. (\ref{numericalmethod}) we obtain the homogeneous linear second-order differential equation for the function $\chi (z)$
\begin{equation}
\chi ^{\prime \prime }=\lambda _{0}(z)\chi ^{\prime }+s_{0}(z)\chi ~,
\label{de}
\end{equation}%
where
\begin{eqnarray}
\notag \lambda _{0}(u)&=&-\frac{r_H^{-z}}{A_1(-1 + u) u f(u)}(\sqrt{(1+A_2)^2 - 4 B_2}A_1 r_H^z u f(u) - 2i l^{(1 + z)} (u-1) \omega f(u)\\
 && + A_1 r_H^z u ((A_2 - z) f(u) + (u-1) f^{\prime}(u)))~,
\end{eqnarray}
\begin{eqnarray}
\notag s_{0}(z) &=&-\frac{r_H^{-2(1+z)}}{2 A_1^2 (u-1)^2 u^2 f(u)^2} (2 A_1^2 l^{2 + 2 z} r_H^2 (1 - u)^{2z} u^2 \omega^2-2 l^{2 + 2 z} (u-1)^2 \omega^2\\
\notag && - 2 i A_1 l^{1 + z} r_H^
    z (u-1) \omega (1 +
      u (-1 + A_2 + \sqrt{(1 + A_2)^2 - 4 B_2} - z))\\ 
  \notag &&-1-\sqrt{(1 + A_2)^2-4B_2}-2B_2A_2+\sqrt{(1+A_2)^2-4B_2}(A_2-z)-z)f(u)^2\\
\notag &&+2 A_1 l^2 r_H^z (-m^2 r_H^2 + Q (u-1)^2) u \\
 &&+(r_H^2 (u-1) ((1+A_2 +\sqrt{(1+A_2)^2-4B_2}) A_1 r_H^z u -
      2 i l^{1+z} (u-1) \omega) f^{\prime}(u)))~,
\end{eqnarray}
where
\begin{equation}
A_1=f^{\prime}(0)~, A_2=1+z~, B_2 = -m^2l^2~.
\end{equation}
That can be solved numerically (see \cite{Becar:2015gca} for more details).
So, we choose the following parameters $l=1$, $m=1$, $q_1=0.01$ and $b=1$. 
Then, in Table \ref{QNM1}, we show the fundamental quasionormal frequency and the first overtone for a massive scalar field $m_s=0.1$ and for a massless scalar field $m_s=0$ with $z=2$, $p=2$ and different values of the momentum angular $Q$. We can observed that the modes are non-overdamped, at zero momenta. Otherwise, the system is always overdamped. 
Then, in Table \ref{QNM2}, we set $Q=0$ and we show some lowest QNFs, for $p=2$, and different values of $z$ for a massive scalar field $m_s=0.1$ and for a massless scalar field $m_s=0$. We observe that there is a limit on the dynamical exponent $z$ ($z\approx 2.3$) above which the system is always overdamped.  
Additionally, in Table \ref{QNM3}  we show some fundamentals QNFs, for $z=2$, $Q=0$ and different values of  the nonlinear exponent $p$ for a massive scalar field $m_s=0.1$, and  for a massless scalar field $m_s=0$, where we can observe that the behavior of the modes (overdamped or non-overdamped) do not depend on $p$. It is worth mentioning that in all the cases analyzed, we observe that the modes have a negative imaginary part, which ensures the stability of four dimensional topological nonlinear charged Lifshitz black holes with spherical transverse section under scalar perturbations.
\begin{table}[ht]
\caption{QNFs  for a massive scalar field $m_s=0.1$ and for a massless scalar field $m_s=0$ with $l=1$, $m=1$, $q_1=0.1$, $b=1$, $z=2$, $p=2$ and different values of $Q$.}
\label{QNM1}\centering
\begin{tabular}{ | c | c | c | c | c | c |  }
\hline
\multicolumn{6}{|c|}{$m_s=0.1$} \\ \hline
$n$ & $Q=0$ & $Q=2$ & $Q=6$ & $Q=12$ & $Q=20$ \\[0.5ex] \hline
$0$ & $0.77567-3.99596i$ &  $-2.86912i$ & $-1.37093i$ & $-3.36678i$ & $-1.99983i$ \\
$1$ & $1.10519-7.98646i$ &  $-4.43686i$ & $-4.32506i$ & $-6.05415i$ & $-5.19559i$ \\ \hline
\multicolumn{6}{|c|}{$m_s=0$} \\ \hline
$0$ & $0.77560-3.99350i$ &  $-2.86614i$ & $-1.36758i$ & $-3.36350i$ & $-1.99611i$ \\
$1$ & $1.10505-7.98396i$ &  $-4.43456i$ & $-4.32229i$ & $-6.05129i$ & $-5.19239i$ \\\hline
\end{tabular}
\end{table}

\begin{table}[ht]
\caption{QNFs  for a massive scalar field $m_s=0.1$ and for a massless scalar field $m_s=0$ with $l=1$, $m=1$, $q_1=0.1$, $b=1$, $p=2$, $Q=0$ and different values of $z$.}
\label{QNM2}\centering
\begin{tabular}{ | c | c | c | c | c | c |  }
\hline
\multicolumn{6}{|c|}{$m_s=0.1$} \\ \hline
$n$ & $z=2$ & $z=2.3$ & $z=3$ & $z=4$ & $z=8$ \\[0.5ex] \hline
$0$ & $0.77567-3.99596i$ & $-3.78732i$ & $-3.58853i$ & $-3.87382i$ & $-5.65842i$ \\
$1$ & $1.10519-7.98646i$ & $-4.68822i$ & $-5.75581i$ & $-6.74363i$ & $-10.62950i$ \\\hline
\multicolumn{6}{|c|}{$m_s=0$} \\ \hline
$0$ & $0.77560-3.99350i$ & $-3.78550i$ & $-3.58710i$ & $-3.87273i$ & $-5.65784i$ \\
$1$ & $1.10505-7.98396i$ &  $-4.68582i$ & $-5.75415i$ & $-6.74243i$ & $-10.62890i$ \\\hline
\end{tabular}%
\end{table}
\begin{table}[ht]
\caption{QNFs  for a massive scalar field $m_s=0.1$ and for a massless scalar field $m_s=0$ with $l=1$, $m=1$, $q_1=0.1$, $b=1$, $z=2$, $Q=0$ and different values of $p$.}
\label{QNM3}\centering
\begin{tabular}{ | c | c | c | c |   }
\hline
\multicolumn{4}{|c|}{$m_s=0.1$} \\ \hline
$n$ & $p=1$ & $p=2$ & $p=2.5$ \\[0.5ex] \hline
$0$ & $0.74428-3.98456i$ &  $0.77567-3.99596i$ & $0.77694-3.99763i$ \\
$1$ & $1.02904-7.96141i$ &  $1.10519-7.98646i$ & $1.10760-7.98995i$ \\\hline
\multicolumn{4}{|c|}{$m_s=0$} \\ \hline
$0$ & $0.74424-3.98210i$ &  $0.77560-3.99350i$ & $0.77687-3.99516i$ \\
$1$ & $1.02895-7.95893i$ &  $1.10505-7.98396i$ & $1.10746-7.98745i$ \\\hline
\end{tabular}%
\end{table}
The results obtained previously can be generalized for a plane transverse section. 
The effective potential has a similar behavior on the horizon and asymptotically that the case of spherical transverse section. Now, in Table \ref{QNM1p}, \ref{QNM2p}, \ref{QNM3p} we show the QNFs for some cases analyzed for spherical transverse section. Here, in all the cases analyzed, we observe that the system is overdamped with a negative imaginary part, which ensures the stability of four dimensional topological nonlinear charged Lifshitz black holes with plane transverse section under scalar perturbations.
 
\begin{table}[ht]
\caption{QNFs  for a massive scalar field $m_s=0.1$ and for a massless scalar field $m_s=0$ with $l=1$, $m=1$, $q_1=0.1$, $b=1$, $z=2$, $p=2$ and different values of $Q$. Plane transverse section.}
\label{QNM1p}\centering
\begin{tabular}{ | c | c | c | c | c | c |  }
\hline
\multicolumn{6}{|c|}{$m_s=0.1$} \\ \hline
$n$ & $Q=0$ & $Q=2$ & $Q=6$ & $Q=12$ & $Q=20$ \\[0.5ex] \hline
$0$ & $-3.96288i$ & $-2.69635i$ & $-1.41179i$ & $-3.38114i$ & $-2.12450i$ \\
$1$ & $-4.03773i$ & $-4.57823i$ & $-4.27695i$ & $-5.99855i$ & $-5.19243i$ \\\hline
\multicolumn{6}{|c|}{$m_s=0$} \\ \hline
$0$ & $-3.96040i$ &  $-2.69346i$ & $-1.40842i$ & $-3.37791i$ & $-2.12082i$ \\
$1$ & $-4.03522i$ &  $-4.57576i$ & $-4.27417i$ & $-5.99572i$ & $-5.18929i$ \\\hline
\end{tabular}%
\end{table}
\begin{table}[ht]
\caption{QNFs for a massive scalar field $m_s=0.1$ and for a massless scalar field $m_s=0$ with $l=1$, $m=1$, $q_1=0.1$, $b=1$, $p=2$, $Q=0$ and different values of $z$. Plane transverse section.}
\label{QNM2p}\centering
\begin{tabular}{ | c | c | c | c | c | c |  }
\hline
\multicolumn{5}{|c|}{$m_s=0.1$} \\ \hline
$n$ & $z=2$ & $z=3$ & $z=4$ & $z=8$ \\[0.5ex] \hline
$0$ & $-3.96288i$ & $-3.52783i$ &  $-3.85397i$ & $-5.65638i$ \\
$1$ & $-4.03773i$ &  $-5.75795i$ & $-6.73535i$ & $-10.62790i$ \\\hline
\multicolumn{5}{|c|}{$m_s=0$} \\ \hline
$0$ & $-3.96040i$ & $-3.52634i$ &  $-3.85284i$ & $-5.65580i$ \\
$1$ & $-4.03522i$ &  $-5.75627i$ & $-6.73412i$ & $-10.62730i$ \\\hline
\end{tabular}%
\end{table}
\begin{table}[ht]
\caption{QNFs  for a massive scalar field $m_s=0.1$ and for a massless scalar field $m_s=0$ with $l=1$, $m=1$, $q_1=0.1$, $b=1$, $z=2$, $Q=0$ and different values of $p$. Plane transverse section.}
\label{QNM3p}\centering
\begin{tabular}{ | c | c | c | c | c | c |  }
\hline
\multicolumn{4}{|c|}{$m_s=0.1$} \\ \hline
$n$ & $p=1$ & $p=2$ & $p=2.5$ \\[0.5ex] \hline
$0$ & $-3.81513i$ &  $-3.96288i$ & $-3.98601i$ \\
$1$ & $-4.16136i$ &  $-4.03773i$ & $-4.01793i$ \\\hline
\multicolumn{4}{|c|}{$m_s=0$} \\ \hline
$0$ & $-3.81272i$ &  $-3.96040i$ & $-3.98352i$ \\
$1$ & $-4.15879i$ &  $-4.03522i$ & $-4.01543i$ \\\hline
\end{tabular}%
\end{table}
 
section{Concluding comments}
\label{conclusion}
In this work we have calculated numerically the QNFs of scalar field perturbations for 
four dimensional topological nonlinear charged Lifshitz black holes with a spherical and plane transverse sections.
Then, we have studied the stability of these black holes under massive and massless scalar field perturbations and we have shown that for all the cases analyzed, the modes have a negative imaginary part, which ensures the stability of four dimensional topological nonlinear charged Lifshitz black holes with spherical and plane transverse section under scalar perturbations. Also, it was found that the modes are overdamped depending strongly of the dynamical exponent and the angular momentum of the scalar field for a spherical transverse section. However,  the modes of a four dimensional topological nonlinear charged Lifshitz black holes with a plane transverse section are always overdamped. 

\acknowledgments

This work was funded by Comisi\'{o}n
Nacional de Ciencias y Tecnolog\'{i}a through FONDECYT Grants 11140674 (PAG). P. A. G. acknowledges the hospitality of the Universidad de La Serena and Universidad Cat\'{o}lica de Temuco and R.B. acknowledges the hospitality of the Universidad Diego Portales where part of this work was undertaken.  


\end{document}